\documentclass[11pt,twoside]{article}
\usepackage{asp2010}

\resetcounters

\begin{document}

\newcommand{\intensunits}{\mbox{erg s$^{-1}$ cm$^{-2}$ Hz$^{-1}$}}
\newcommand{\kmsend}{\mbox{km s$^{-1}$}}
 \newcommand{\kms}{\mbox{km s$^{-1}$ }} 
\newcommand{\hgyr}{\mbox{h$^{-1}$Gyr }}
\newcommand{\hmyr}{\mbox{h$^{-1}$Myr }}
\newcommand{\hkpc}{\mbox{h$^{-1}$kpc }}
\newcommand{\hpc}{\mbox{h$^{-1}$pc }}
\newcommand{\hmsun}{\mbox{h$^{-1}$M$_{\sun}$ }}
\newcommand{\hmsunend}{\mbox{h$^{-1}$M$_{\sun}$}}
\newcommand{\msun}{\mbox{M$_{\sun}$ }}
\newcommand{\msunend}{\mbox{M$_{\sun}$}}
\newcommand{\lsun}{\mbox{L$_{\sun}$ }}
\newcommand{\lsunend}{\mbox{L$_{\sun}$}}
\newcommand{\lir}{\mbox{L$_{\rm IR}$}}
\newcommand{\lhcn}{\mbox{L$_{\rm HCN}$}}
\newcommand{\lhcnjone}{\mbox{HCN (J=1-0)}}
\newcommand{\lco}{\mbox{L$_{\rm CO}$}}
\newcommand{\lmolecule}{\mbox{L$_{\rm molecule}$}}
\newcommand{\lmol}{\mbox{L$_{\rm mol}$}}
\newcommand{\ltwentyfive}{\mbox{L$_{\rm 25}$}}
\newcommand{\lcojone}{\mbox{CO (J=1-0)}}
\newcommand{\lcojthree}{\mbox{CO (J=3-2)}}
\newcommand{\ncrit}{\mbox{$n_{\rm crit}$}}

\newcommand{\cmthree}{\mbox{cm$^{-3}$}}
\newcommand{\cmtwo}{\mbox{cm$^{-2}$}}
\newcommand{\msunyr}{\mbox{M$_{\sun}$yr$^{-1}$ }}
\newcommand{\msunyrend}{\mbox{M$_{\sun}$yr$^{-1}$}}
\newcommand{\htwo}{\mbox{H$_2$}}
\newcommand{\mhtwo}{\mbox{$M_{H2}$}}
\newcommand{\z}{\mbox{$z$}}
\newcommand{\zapprox}{\mbox{$z \approx $}}
\newcommand{\zga}{\mbox{$z \ga$}}
\newcommand{\zla}{\mbox{$z \la$}}
\newcommand{\zsim}{\mbox{$z\sim$ }}
\newcommand{\msigma}{\mbox{$M_{\rm BH}$-$\sigma_{\rm v}$ }}
\newcommand{\magorrian}{\mbox{$M_{\rm BH}$-$M_{\rm bulge}$ }}
\newcommand{\lbol}{L$_{\rm {bol}}$}
\newcommand{\MBH}{M$_{\rm{BH}}$}
\newcommand{\Msun}{M$_{\odot}$}
\newcommand{\Lsun}{L$_{\odot}$}

\newcommand{\myr}{\rm {Myr}}
\newcommand{\Mvir}{M_{\rm{vir}}}
\newcommand{\Vvir}{V_{\rm{vir}}}

\newcommand{\arttwo}{\mbox{\it ART}2}
\newcommand{\mdyn}{\mbox{M$_{\rm dyn}$}}
\newcommand{\mactual}{\mbox{M$_{\rm actual}$}}
\newcommand{\sef}{\mbox{$F_{\rm 850}$}}
\newcommand{\stf}{\mbox{$F_{\rm 24}$}}
\newcommand{\sr}{\mbox{$F_{\rm R}$}}
\newcommand{\tfr}{\mbox{24/$R$}}
\newcommand{\microjy}{\mbox{$\mu$Jy}}
\newcommand{\sunrise}{\mbox{\sc sunrise}}
\newcommand{\gadget}{\mbox{\sc gadget-3}}
\newcommand{\gadgettwo}{\mbox{\sc gadget-2}}
\newcommand{\starburst}{\mbox{\sc starburst99}}
\newcommand{\mappings}{\mbox{\sc mappingsiii}}
\newcommand{\turtlebeach}{\mbox{\sc turtlebeach}}
\newcommand{\bzk}{\mbox{{\it BzK}}}

\newcommand{\schmidt}{\mbox{SFR $\propto \rho^N$}}
\newcommand{\sigmacojone}{\mbox{$\Sigma_{\rm SFR}-\Sigma_{\rm CO J=1-0}^\alpha$}}
\newcommand{\sigmacojthree}{\mbox{$\Sigma_{\rm SFR}-\Sigma_{\rm CO J=3-2}^\alpha$}}
\newcommand{\sigmaco}{\mbox{$\Sigma_{\rm SFR}-\Sigma_{\rm CO}^\alpha$}}
\newcommand{\sigmahtwo}{\mbox{$\Sigma_{\rm SFR}-\Sigma_{\rm H2}^\alpha$}}
\newcommand{\sigmamol}{\mbox{$\Sigma_{\rm SFR}-\Sigma_{\rm mol}^\alpha$}}
\newcommand{\nhtwo}{\mbox{$N_{\rm H2}$}}
\newcommand{\xco}{\mbox{$X_{\rm CO}$}}
\newcommand{\alphaco}{\mbox{$\alpha_{\rm CO}$}}
\newcommand{\xcounits}{\mbox{cm$^{-2}$/K-km s$^{-1}$}}
\newcommand{\alphacounits}{\mbox{\msun pc$^{-2}$ (K-\kmsend)$^{-1}$}}
\newcommand{\qeos}{\mbox{$q_{\rm EOS}$}}

\newcommand{\fgas}{\mbox{$f_{\rm gas}$}}

\title{The CO-\htwo \ Conversion Factor in Galaxy Mergers
\author{Desika Narayanan$^1,^2$
\affil{$^1$Steward Observatory, University of Arizona \\ $^2$Bart J Bok Fellow}
}}

\begin{abstract}
The CO-\htwo \ conversion factor in galaxies is typically described as
bimodal: one value for discs and quiescent regions, and another
(lower) value for mergers and starbursts.  In this proceeding, I will
describe both empirical observational evidence that the conversion
factor varies with physical environment, as well as a theoretical
model which aims to understand the physical processes which drive
these variations.  I present a functional form for \xco \ which can be
applied to observations ranging in scale from $\sim 70$ pc to
galaxy-wide scales, and show the consequences of the application of
this model to the Kennicutt-Schmidt star formation law.
\end{abstract}

\section{Introduction}

Deriving an \htwo \ gas mass from a giant molecular cloud (GMC) or
galaxy full of clouds involves the usage of tracer molecules such as
$^{12}$CO (hereafter, CO).  This is because \htwo \ has no permanent
dipole moment, and its first quadrupole moment lies $\sim 500$ K above
ground, significantly warmer than the typical ISM temperature in a
GMC.  Converting from a CO emission line strength to an underlying
\htwo \ gas mass  involves the usage of a CO-\htwo
\ conversion factor.

The CO-\htwo \ conversion factor goes by two names in the literature:
\xco\footnote{\xco \ is also referred to as the $X$-factor.  I will
  use \xco \ or the $X$-factor arbitrarily in this proceeding.} \ and
\alphaco. The former has units of \xcounits, while the latter \msun
${\rm pc}^{-2}$(K-\kmsend)$^{-1}$.  The two are equivalent, and simply
related via \xco = $6.3 \times 10^{19} \alpha_{\rm CO}$ in the
aforementioned units.  Since both \xco \ and \alphaco \ are used in
the literature, I will present the results of this proceeding in terms
of both units.

Observationally determining an $X$-factor requires deriving
an independent measurement of \htwo \ gas mass, and comparing this 
to an observed CO line flux.  Via a variety of methods \citep[some of
  these are summarised in the seminal conference proceeding
  by][]{sol91}, it has been found that the conversion factor is
roughly constant in the Galaxy and nearby galaxies (when the
metallicity is of order solar) with value \xco= $2-4 \times 10^{20}
\xcounits$, or \alphaco $\approx 3-6$ \alphacounits \ \citep[for a
  reference list of these observations, please see the introduction of][]{nar11c}.

\begin{figure}
\includegraphics[scale=0.4,angle=90]{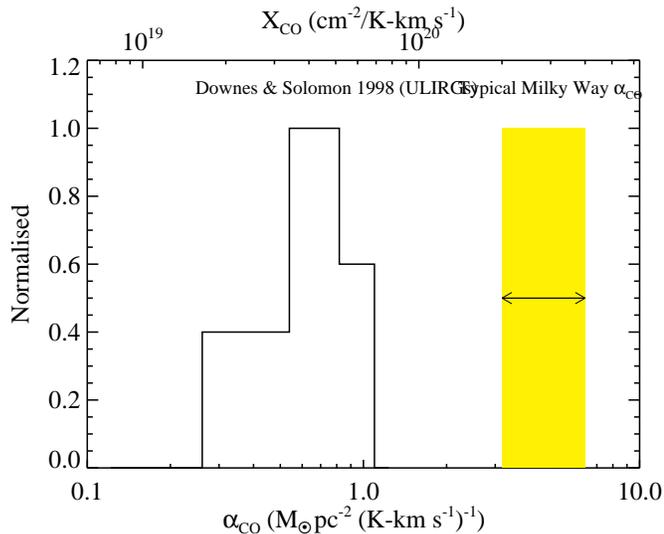}
\caption{Histogram of the CO-\htwo \ conversion factors measured for
  local ULIRGs by \citet{dow98} (solid black line).  The yellow shaded
  region denotes the typical values found in the Milky Way.  Despite a
dispersion of conversion factor values for mergers, a singular value
  of \alphaco=0.8 is typically assumed for merging
  galaxies.  \label{figure:observational_xco}}
\vspace{-.15in}
\end{figure}

However, despite the seeming constancy of the $X$-factor in relatively
``normal'' GMCs, observations of nearby ultraluminous infrared
galaxies (ULIRGs) show that the usage of a Milky Way \xco \ in these
environments would cause the inferred \htwo \ gas mass to exceed the
measured dynamical mass \citep{dow98}.  Hence, in ULIRGs, the
$X$-factor must be lower than within the Galaxy.  In
Figure~\ref{figure:observational_xco}, we show the distribution of
\xco \ and \alphaco \ values from the \citet{dow98} survey, as well as
a shaded region for the typical range of Galactic values.  There is a
dispersion amongst the ULIRG $X$-factors, though on average they are
lower than the Galactic mean.  Despite the observed dispersion in
ULIRG conversion factors, the fact that they are lower on average than
the Milky Way value has led the community to largely adopt a bimodal
picture of \xco: a value of $\alphaco \approx 4$ for disc galaxies,
and $\alphaco \approx 0.8$ for starbursts and mergers.

 The ramifications of a bimodal form of the CO-\htwo \ conversion
 factor are significant.  As an example, \citet{dad10b} and
 \citet{gen10} demonstrated the effects of assuming a bimodal \xco
 \ on the Kennicutt-Schmidt (KS) star formation rate-gas density
 relation.  While in $\Sigma_{\rm SFR}-\Sigma_{\rm CO}$ space,
 galaxies at low and high-\z \ lie on an arguably unimodal relation,
 the introduction of a bimodal \xco \ to convert the CO luminosity to
 an \htwo \ gas mass causes the KS relation to become bimodal.  That
 is, mergers and discs lie on different track on the KS relation when
 utilising a different \xco \ for each.  This is shown explicitly in
 Figure~\ref{figure:ks}, where I have compiled the galaxies presented
 in \citet{dad10b} and \citet{gen10}, and plotted (in the left panel)
 the $\Sigma_{\rm SFR}-\Sigma_{\rm CO}$ relation, and (in the middle
 panel), the $\Sigma_{\rm SFR}-\Sigma_{\rm H2}$ relation which
 utilises a bimodal $X$-factor.  The mergers (squares) and discs
 (circles) occupy different tracks.  This has given rise to a
 terminology in the literature, as well as at conferences (including
 this one!), that mergers and discs have different ``modes'' of star
 formation.  Mergers, according to this relation, form stars more
 efficiently (i.e. needing less gas to sustain a given SFR) than
 discs.  We will neglect the right panel of Figure~\ref{figure:ks} for
 now, and will return to it shortly.

\section{The Case Against a Bimodal Conversion Factor: Empirical Observational Evidence}
\label{section:caseagainst}

The usage of a bimodal CO-\htwo \ conversion factor raises a number of
difficult questions. For example, if one ascribes a mean
``merger/starburst'' value of \xco \ to all mergers, should this value
vary with the interaction stage of a merger?  What about with close
pairs which do not have overlapping discs?  Does the same
``merger/starburst'' value apply for mergers of all gas fractions,
orientation angles, star formation rates, mass ratios and galaxy
masses?  Does this merger value evolve with redshift, or should one
utilise the locally-calibrated $X$-factor for mergers at high-\z
\ (which are presumably more gas rich and have higher star formation
rates than their local counterparts)?  Which value should one use for
\zsim 2 discs, which are forming stars at rates comparable to local
mergers?

An alternative picture to the bimodal $X$-factor is one in which the
CO-\htwo \ conversion factor varies with the physical conditions in
the ISM.  Intuitively, this scenario makes sense.  The fraction of
hydrogen in molecular form, fraction of carbon in the form of CO, and
escape of CO radiation from GMCs is all dependent on the physical
conditions of the ISM.  So, why shouldn't the CO-\htwo \ conversion
factor also depend on the physical conditions?  In fact, there is
substantial empirical empirical evidence that it does.

 First, as is shown in the Appendix of \citet{tac08}, \xco \ is seen
 to vary smoothly with the \htwo \ surface density of galaxies.  This
 was found by \citet{ost11} to be well fit by a relation \alphaco
 $\sim \Sigma_{\rm H2}^{-0.52} \sim W_{\rm CO}^{-0.34}$ where
 $\Sigma_{\rm H2}$ is the surface density of molecular gas, and
 $W_{\rm CO}$ is the observed CO intensity (in K-\kmsend).

Second, the $X$-factor appears to vary with the metallicity of the
gas.  Recent work by \citet{bol08}, \citet{ler11} and \citet{gen11b} has shown
that \xco \ increases with metallicity with power $X \propto$
(O/H)$^{-b}$ where $b=1-2.7$ \citep{ari96,isr97}.  The physical
motivation behind this sort of scaling is that in low metallicity
clouds, \htwo \ can self-shield for survival from photodissociating
radiation whereas CO requires some amount of dust to protect
it. Hence, the fraction of CO-dark \htwo \ mass rises as metallicity
decreases \citep{wol10}.  

Right away, the results from \citet{ost11} alongside the
aforementioned results regarding the variation of \xco \ with
metallicity imply an empirical relation which looks like:
\begin{equation}
\xco \propto \alphaco \propto W_{\rm CO}^{-\beta}Z^{-b}
\end{equation}

\section{A Theory for the CO-\htwo \ Conversion Factor}

\subsection{Numerical Models}
We will utilise numerical simulations to try to understand the origins
of variations in the CO-\htwo \ conversion factor.  The principle
equations for this model are described in \citet{nar11b}, and I
encourage the reader to refer to this paper.  The description here
will be necessarily abbreviated owing to space limitations.  We first
model the evolution of idealised galaxies hydrodynamically utilising
the publicly available SPH code \gadget \ \citep{spr05b}.  We model
isolated disc galaxies, and galaxy mergers over a range of masses,
merger mass ratios, and redshifts.  We smooth the physical conditions
onto an adaptive mesh for the purposes of radiative transfer.

\begin{figure}
\hspace{-1in}
\includegraphics[angle=90,scale=0.75]{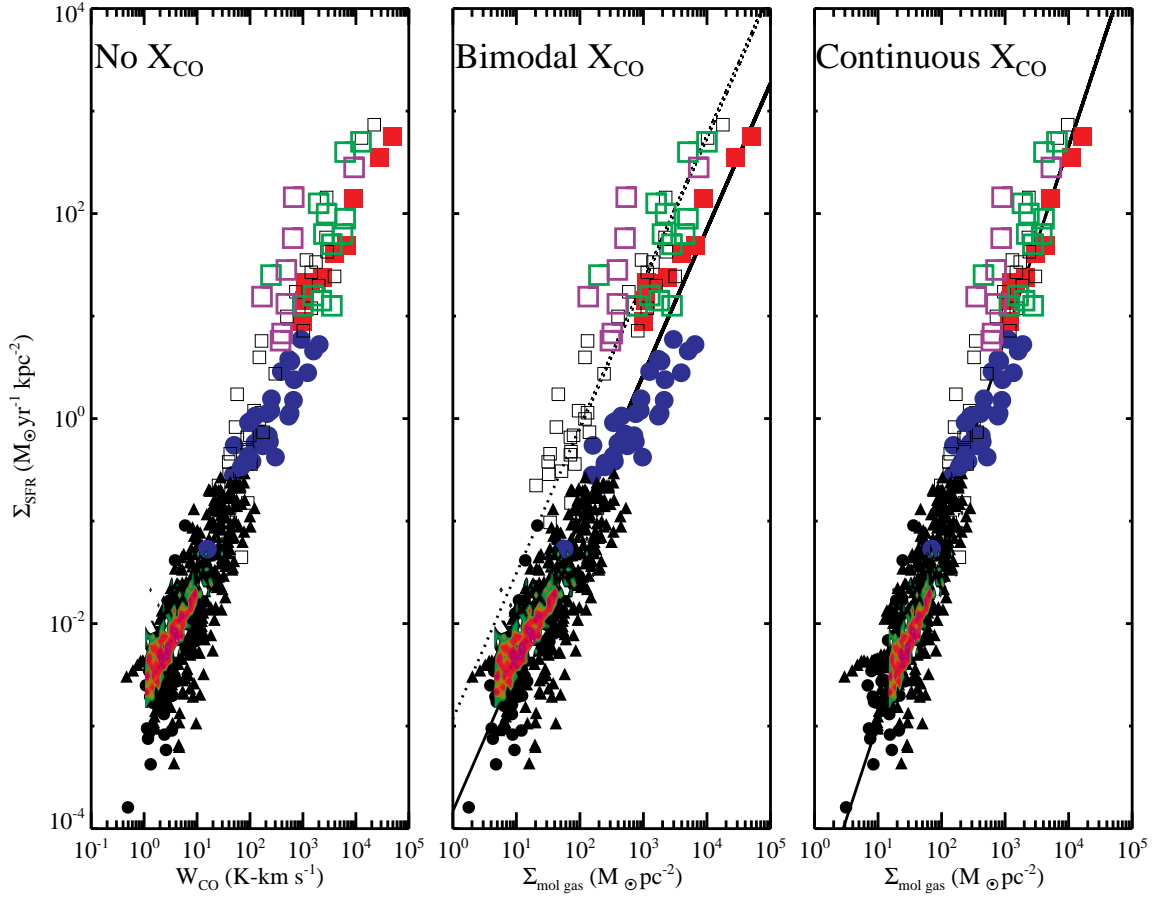}
\caption{Kennicutt-Schmidt star formation relation in observed
  galaxies.  Circles and triangles are local discs or high-\z \ \bzk
  \ galaxies (inferred high-\z \ discs), and squares are inferred
  mergers (local ULIRGs or high-\z \ SMGs).  Colours denote surveys
  galaxies are pulled from and are described in \citet{nar11c}.  {\it
    Left}: SFR surface density vs. velocity-integrated CO intensity.
  In terms of the observable, $W_{\rm CO}$, a unimodal SFR relation
  exists.  {\it Centre}: When applying an effectively bimodal \xco
  \ ($\alpha_{\rm CO}= 4.5$ for local discs, 3.6 for high-\z \ discs,
  and 0.8 for mergers), the resulting SFR relation is bimodal, with
  mergers occupying a track of higher star formation efficiency
  (defined as SFR/M$_{\rm H2}$).  The solid and dotted lines
  overplotted are the best fit tracks for each ``mode'' of star
  formation as published in \citet{dad10b}. {\it Right}: Resulting SFR
  relation when applying Equation~\ref{equation:xco} to the data.
  When \xco \ varies smoothly with galactic physical properties, the
  result is a unimodal SFR relation.  In this case the power-law index
  in the relation is approximately 2 (when assuming that each galaxy
  has metallicity $Z'=1$). \label{figure:ks}\label{figure:ks}}
\end{figure}

The \htwo \ fraction is calculated following the prescriptions of
\citet{kru08,kru09b} which balances the photodissocation by
Lyman-Werner band photons against growth on dust grains.  We assume
that all \htwo \ gas is assumed to be in spherical GMC.  Because some
cells (which contain little \htwo \ mass) can be rather large in the
adaptive mesh, we establish a floor surface density of 100 \msun
pc$^{-2}$.  In our model mergers, the bulk of the molecular mass is in
GMCs which are above this floor value.  We move beyond the
10,000 K floor temperature established in \gadget \ by calculating the
temperature of the molecular gas as a balance between line cooling,
heating by the grain photoelectric effect, cosmic rays, and energy
exchange with dust \citep{kru11a}.

To calculate the CO emission line fluxes, we first determine the
emergent radiation from each GMC utilising the publicly available
escape probability code of \citet{kru07}.  We then run the 3D Monte
Carlo radiative transfer package \turtlebeach \ \citep{nar08a_let}
to calculate the radiative transfer through the galaxy.  With
modeled CO line fluxes, and knowledge of the \htwo \ content in
our model galaxies, we are prepared to study the CO-\htwo \ conversion
factor in our simulations.  

\subsection{Results from Simulations}

At a given metallicity, \xco \ decreases with increasing gas temperature
and velocity dispersion.  The reason for this is straightforward.  The
velocity integrated CO line intensity enters \xco \ in the
denominator.  As gas temperature increases, the peak intensity of the CO
emission line, which is in the Rayleigh-Jeans limit, also increases.
Similarly, because GMCs are typically optically thick, as the gas
velocity increases, the velocity integrated CO line intensity
increases.

This is the reason \xco \ is lower in galaxy mergers than the Galactic
mean value \citep{nar11b}.  In a galaxy merger, the velocity
dispersion of the model GMCs in our simulations increase from the
virial velocity of the GMCs due to the turbulent nature of the gas in
a galaxy merger.  The velocity dispersion within a GMC in the merger
can be as large as $50-100$ \kms (compared to a few \kms in a
quiescent cloud).  Similarly, the gas temperature goes up in
GMCs. While the typical temperature of a quiescent GMC is of order
$\sim 10$ K, representative of the floor temperature established by a
Galactic cosmic ray heating rate, the temperatures in GMCs in galaxy
mergers can be of order $\sim 50-100$ K.  The reason for this is due
to energy exchange with dust.  When gas densities rise above $\sim
10^{4}$ \cmthree, dust and gas exchange energy extremely efficiently,
and the gas temperature rises to the dust temperature \citep{gol01}.
In a merger, not only are these large gas densities typical in GMCs,
the dust temperature is elevated due to the increased star formation
rate \citep[see Figure 2 of][]{nar11b}. A consequence of the increased
gas temperatures and velocity dispersions in galaxy mergers is a
decreased \xco.  This may explain the results of \citet{dow98} and
Figure~\ref{figure:observational_xco}.

A second physical effect controlling the conversion factor is the gas
phase metallicity.  As explained before, in low-metallicity gas, the
carbon resides predominantly in the form of CI or CII (rather than CO)
due to photodissocation, whereas \htwo \ can self-shield to survive.
In this regime, the velocity-integrated CO intensity decreases for a
given \htwo \ gas mass, and the $X$-factor rises.  MHD models of GMCs
by \citet{glo11} and \citet{she11a,she11b}, as well as cosmological
galaxy evolution simulations (R. Feldmann et al., submitted) find
similar trends.

While our model isolated discs have mean $X$-factors comparable to the
Milky Way, and our model \z=0 major mergers have $X$-factors
consistent with the observed range presented in
Figure~\ref{figure:observational_xco}, there is no ``disc'' value and
``merger'' value of \xco.  In a minor merger, for example, the rise in
gas temperatures and velocity dispersions are not as extreme as they
are in 1:1 major merger.  Hence, the rise in velocity-integrated CO
intensity is not as large, and the $X$-factor does not decrease as
much.  Similarly, in high-\z \ disc galaxies, large gas clumps that
are forming stars rapidly due to gravitational instabilities in a
gas-rich environment have larger temperatures and velocity dispersions
than quiescent \z=0 discs.  Hence, they have $X$-factors lower than
the locally calibrated ``quiescent/disc'' value.  There is no
``merger'' value and ``disc'' value of \xco \ \citep{nar11c}.

The good news is that the CO-\htwo \ conversion factor can be
observationally calibrated.  While the gas temperature and velocity
dispersion are not easily observable, in our models, the CO surface
brightness (line intensity) serves as a good proxy for the product of
these quantities.  That is, larger CO line intensities tend to
correlate with larger velocity dispersions and temperatures.  Fitting
\xco \ in our simulations against the CO line intensity, $W_{\rm CO}$,
and metallicity, we get:
\begin{eqnarray}
\label{equation:xco}
\xco = \frac{6.75 \times 10^{20} \times \langle W_{\rm CO}\rangle^{-0.32}}{Z'^{0.65}}\\
\alphaco = \frac{10.7 \times \langle W_{\rm CO} \rangle^{-0.32}}{Z'^{0.65}}
\end{eqnarray}
where $\langle W_{\rm CO} \rangle$ is the luminosity-weighted CO
intensity over all GMCs in a galaxy, and $Z'$ is the metallicity in
units of solar. In the limit of a uniform distribution of luminosity
from the ISM in a galaxy, $\langle W_{\rm CO} \rangle$ reduces to
$L'_{\rm CO}/A$, or the CO intensity ($A$ is the area observed). Note
the similarity of our model fit for \xco, and what was derived from
empirical observations in \S~\ref{section:caseagainst}.  Hence, with a
metallicity estimate and a CO surface brightness, one can calculate
the $X$-factor observationally.  I note that there is an implicit
ceiling value of \xco = $4 \times 10^{20}$ \xcounits (\alphaco = 6.3
\alphacounits) when $Z'=1$ in Equations 2 and 3.

\section{The Ramifications of This Model Form for \xco}
\subsection{The Kennicutt-Schmidt Star Formation Relation}
The usage of Equations 2 and 3 has an immediate consequence
for the observed Kennicutt-Schmidt star formation relation.  Recall
that when a bimodal conversion factor is used for mergers and discs, a
bimodal KS relation results with different normalisations for mergers
and discs (centre panel of Figure~\ref{figure:ks}.  When utilising our
continuous form for \xco, this bimodality goes away.  

We now turn to the right panel of Figure~\ref{figure:ks}, where we
have applied our model form for \xco \ to the observed data in the
\citet{dad10b} and \citet{gen10} compilation.  The bimodal
relationship is reduced to a unimodal one with best fit index is a
relationship $\Sigma_{\rm SFR} \sim \Sigma_{\rm mol}^2$. In this
picture, there are no different ``modes'' of star formation in mergers
and discs.  On {\it average}, mergers do form stars more efficiently
than disc galaxies as they lie preferentially toward the high
$\Sigma_{\rm SFR}$ regime of Figure~\ref{figure:ks}.  However, if one
examines the central region of the fit, where both high-\z \ discs
(filled blue circles) and local galaxy mergers (open black squares)
reside, it is evident that they have similar star formation
efficiencies (defined as SFR/M$_{\rm H2}$).  A key point of this model
is that {\it if the physical conditions in a disc and a merger are the
  same, they will have the same conversion factor.}

\subsection{The Typical Conversion Factors for Different Galaxy Populations}

\begin{figure}
\includegraphics[angle=90,scale=0.5]{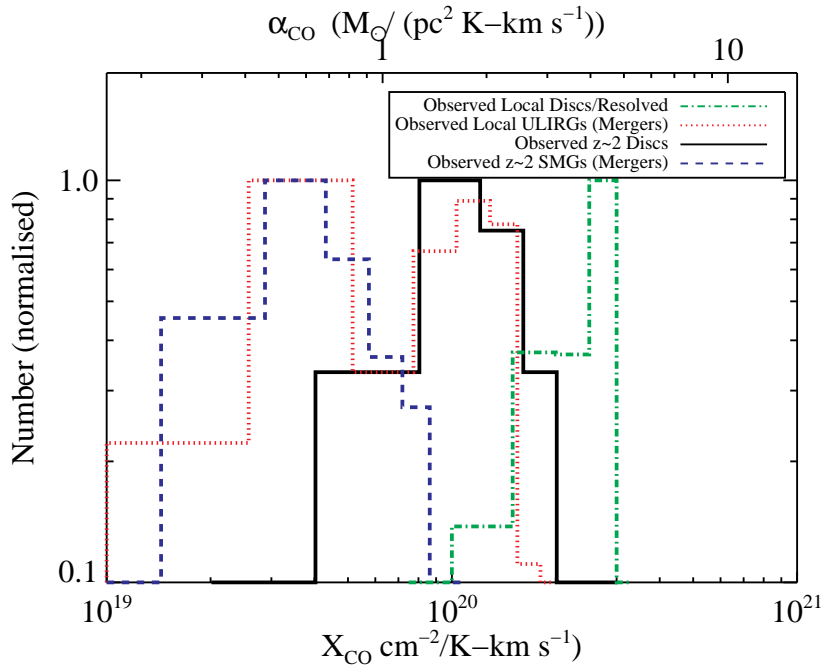}
\caption{Histograms of the derived conversion factors for all the
  galaxies in Figure~\ref{figure:ks}.\label{figure:xco_realgalaxies}}
\end{figure}

This point is made more clearly in
Figure~\ref{figure:xco_realgalaxies}, where we show the conversion
factors we derive for the galaxies in Figure~\ref{figure:ks} utilising
Equations~\ref{equation:xco}.  Here, we see that {\it on average}
mergers have lower conversion factors than discs.  The high-\z \ SMGs
(blue dashed line) have the lowest \xco \ values, followed by low-\z
\ mergers (red dotted line).  However, there is substantial overlap in
the conversion factors, especially between the low-\z \ mergers and
high-\z \ discs (black solid line).

This model clearly predict, then, that attempts to determine
conversion factors for high-\z \ discs will evidence a lower average
value than the Galactic mean value.  A few attempts at constraining
\xco \ in high-\z \ \bzk \ galaxies exist.  \citet{dad10a} developed
dynamical models for two high-\z \ \bzk \ galaxies.  After subtracting
off the measured stellar and assumed dark matter masses, they were
able to derive an \xco \ factor by relating the remaining
(presumably) \htwo \ mass to observed CO luminosity.  This method
recovered a mean $X$-factor of $\sim 2 \times 10^{20}$ \xcounits, or
$\alphaco \approx 3 \ \alphacounits$.   \citet{mag11}
recovered similar results via dust to gas ratio arguments.  Our model
predicts that further measurements of the conversion factor from
high-\z \ discs will find a mean value a factor of a few lower than
this (Figure~\ref{figure:xco_realgalaxies}).

 \section{Conclusions}

The principal result from this work is that the CO-\htwo \ conversion
factor in galaxies varies with the physical conditions in the galaxy.
It is not  bimodal with the global
morphology of the galaxy.  Rather, the physical environment sets the CO-\htwo
\ conversion factor.  

The dominant drivers in the conversion factor are the gas temperature,
velocity dispersion and metallicity.  While the former two are
difficult to observe, they can be parameterised by the observable CO
intensity.  With this, the CO-\htwo \ conversion factor can be
described as function of CO intensity and gas phase metallicity.  We
show that this function can be derived both from empirical
observational results, as well as numerical modeling.


\acknowledgements I would like to thank the organizers of the {\it
  Galaxy Mergers in an Evolving Universe} for inviting me to such an
educational meeting.  Thanks also to Nick Scoville for setting the
tone early in the meeting that there should be no limit to the number
of questions.  It was fun getting to field a barrage of questions
(mostly from Nick!) for 20 minutes after my talk.  Thanks go to Mark
Krumholz, Eve Ostriker, and Lars Hernquist who I have collaborated
extensively with in developing this model.  Finally, thanks to the NSF
for funding this work through grant AST-1009452.


\begin{thebibliography}{}
\expandafter\ifx\csname natexlab\endcsname\relax\def\natexlab#1{#1}\fi
\expandafter\ifx\csname url\endcsname\relax
  \def\url#1{\texttt{#1}}\fi
\expandafter\ifx\csname urlprefix\endcsname\relax\def\urlprefix{URL }\fi
\providecommand{\eprint}[2][]{\url{#2}}

\end{thebibliography}


\begin{thebibliography}{}
\expandafter\ifx\csname natexlab\endcsname\relax\def\natexlab#1{#1}\fi
\expandafter\ifx\csname url\endcsname\relax
  \def\url#1{\texttt{#1}}\fi
\expandafter\ifx\csname urlprefix\endcsname\relax\def\urlprefix{URL }\fi
\providecommand{\eprint}[2][]{\url{#2}}

\bibitem[{{Arimoto} et~al.(1996){Arimoto}, {Sofue}, \& {Tsujimoto}}]{ari96}
{Arimoto}, N., {Sofue}, Y., \& {Tsujimoto}, T. 1996, PASJ, 48, 275

\bibitem[{{Bolatto} et~al.(2008){Bolatto}, {Leroy}, {Rosolowsky}, {Walter}, \&
  {Blitz}}]{bol08}
{Bolatto}, A.~D., {Leroy}, A.~K., {Rosolowsky}, E., {Walter}, F., \& {Blitz},
  L. 2008, \apj, 686, 948. \eprint{0807.0009}

\bibitem[{{Daddi} et~al.(2010{\natexlab{a}})}]{dad10b}
{Daddi}, E., et~al. 2010{\natexlab{a}}, \apjl, 714, L118. \eprint{1003.3889}

\bibitem[{{Daddi} et~al.(2010{\natexlab{b}})}]{dad10a}
--- 2010{\natexlab{b}}, \apj, 713, 686. \eprint{0911.2776}

\bibitem[{{Downes} \& {Solomon}(1998)}]{dow98}
{Downes}, D., \& {Solomon}, P.~M. 1998, \apj, 507, 615.
  \eprint{arXiv:astro-ph/9806377}

\bibitem[{{Genzel} et~al.(2010)}]{gen10}
{Genzel}, R., et~al. 2010, \mnras, 407, 2091. \eprint{1003.5180}

\bibitem[{{Genzel} et~al.(2011)}]{gen11b}
--- 2011, ArXiv e-prints. \eprint{1106.2098}

\bibitem[{{Glover} \& {Mac Low}(2011)}]{glo11}
{Glover}, S.~C.~O., \& {Mac Low}, M.-M. 2011, \mnras, 412, 337.
  \eprint{1003.1340}

\bibitem[{{Goldsmith}(2001)}]{gol01}
{Goldsmith}, P.~F. 2001, \apj, 557, 736

\bibitem[{{Israel}(1997)}]{isr97}
{Israel}, F.~P. 1997, \aap, 328, 471. \eprint{arXiv:astro-ph/9709194}

\bibitem[{{Krumholz} et~al.(2011){Krumholz}, {Leroy}, \& {McKee}}]{kru11a}
{Krumholz}, M.~R., {Leroy}, A.~K., \& {McKee}, C.~F. 2011, \apj, 731, 25.
  \eprint{1101.1296}

\bibitem[{{Krumholz} et~al.(2008){Krumholz}, {McKee}, \& {Tumlinson}}]{kru08}
{Krumholz}, M.~R., {McKee}, C.~F., \& {Tumlinson}, J. 2008, \apj, 689, 865.
  \eprint{0805.2947}

\bibitem[{{Krumholz} et~al.(2009){Krumholz}, {McKee}, \& {Tumlinson}}]{kru09b}
--- 2009, \apj, 699, 850. \eprint{0904.0009}

\bibitem[{{Krumholz} \& {Thompson}(2007)}]{kru07}
{Krumholz}, M.~R., \& {Thompson}, T.~A. 2007, \apj, 669, 289.
  \eprint{arXiv:0704.0792}

\bibitem[{{Leroy} et~al.(2011)}]{ler11}
{Leroy}, A.~K., et~al. 2011, arXiv/1102.4618. \eprint{1102.4618}

\bibitem[{{Magdis} et~al.(2011)}]{mag11}
{Magdis}, G.~E., et~al. 2011, arXiv/1109.1140. \eprint{1109.1140}

\bibitem[{{Narayanan} et~al.(2011{\natexlab{a}}){Narayanan}, {Krumholz},
  {Ostriker}, \& {Hernquist}}]{nar11b}
{Narayanan}, D., {Krumholz}, M., {Ostriker}, E.~C., \& {Hernquist}, L.
  2011{\natexlab{a}}, \mnras, 418, 664. \eprint{1104.4118}

\bibitem[{{Narayanan} et~al.(2011{\natexlab{b}}){Narayanan}, {Krumholz},
  {Ostriker}, \& {Hernquist}}]{nar11c}
{Narayanan}, D., {Krumholz}, M.~R., {Ostriker}, E.~C., \& {Hernquist}, L.
  2011{\natexlab{b}}, arXiv/1110.3791. \eprint{1110.3791}

\bibitem[{{Narayanan} et~al.(2008)}]{nar08a_let}
{Narayanan}, D., et~al. 2008, \apjs, 176, 331. \eprint{arXiv:0710.0384}

\bibitem[{{Ostriker} \& {Shetty}(2011)}]{ost11}
{Ostriker}, E.~C., \& {Shetty}, R. 2011, \apj, 731, 41. \eprint{1102.1446}

\bibitem[{{Shetty} et~al.(2011{\natexlab{a}})}]{she11b}
{Shetty}, R., et~al. 2011{\natexlab{a}}, arXiv/1104.3695. \eprint{1104.3695}

\bibitem[{{Shetty} et~al.(2011{\natexlab{b}})}]{she11a}
--- 2011{\natexlab{b}}, \mnras, 412, 1686. \eprint{1011.2019}

\bibitem[{{Solomon} \& {Barrett}(1991)}]{sol91}
{Solomon}, P.~M., \& {Barrett}, J.~W. 1991, in Dynamics of Galaxies and Their
  Molecular Cloud Distributions, edited by F.~{Combes}, \& F.~{Casoli}, vol.
  146 of IAU Symposium, 235

\bibitem[{{Springel}(2005)}]{spr05b}
{Springel}, V. 2005, \mnras, 364, 1105. \eprint{arXiv:astro-ph/0505010}

\bibitem[{{Tacconi} et~al.(2008)}]{tac08}
{Tacconi}, L.~J., et~al. 2008, \apj, 680, 246. \eprint{0801.3650}

\bibitem[{{Wolfire} et~al.(2010){Wolfire}, {Hollenbach}, \& {McKee}}]{wol10}
{Wolfire}, M.~G., {Hollenbach}, D., \& {McKee}, C.~F. 2010, \apj, 716, 1191.
  \eprint{1004.5401}

\end{thebibliography}
\end{document}